\documentclass[12pt]{iopart}

\usepackage{amsmath}
\usepackage{graphicx}
\usepackage{subcaption}
\usepackage{multirow}
\usepackage{url}
\usepackage{float}
\usepackage{array}
\usepackage{bm}

\begin{document}

\title{D-SELD: \textbf{D}ataset-\textbf{S}calable \textbf{E}xemplar \textbf{L}CA-\textbf{D}ecoder}

\author{Sanaz M. Takaghaj, Jack Sampson\\
 }

\vspace{10pt}

\begin{abstract}
Neuromorphic computing has recently gained significant attention as a promising approach for developing energy-efficient, massively parallel computing systems inspired by the spiking behavior of the human brain and natively mapping Spiking Neural Networks (SNNs). Effective training algorithms for SNNs are imperative for increased adoption of neuromorphic platforms; however, SNN training continues to lag behind advances in other classes of ANN. In this paper, we reduce this gap by proposing an innovative encoder-decoder technique that leverages sparse coding and the Locally Competitive Algorithm (LCA) to provide an algorithm specifically designed for neuromorphic platforms. Using our proposed Dataset-Scalable Exemplar LCA-Decoder we reduce the computational demands and memory requirements associated with training SNNs using error backpropagation methods on increasingly larger training sets. We offer a solution that can be scalably applied to datasets of any size. Our results show the highest reported top-1 test accuracy using SNNs on the ImageNet and CIFAR100 datasets, surpassing previous benchmarks. Specifically, we achieved a record top-1 accuracy of 80.75\% on ImageNet (ILSVRC2012 validation set) and 79.32\% on CIFAR100 using SNNs. 
\end{abstract}

\section{Introduction}
\label{sec:introduction}
Neuromorphic computing, inspired by the capabilities of the human brain, aims to revolutionize the field of computing by offering energy-efficient and massively parallel processing elements. This paradigm holds great promise in addressing the limitations of conventional computing systems, such as data movement, power consumption, and real-time processing of complex data, especially for artificial intelligence applications. Conventional processors, like CPUs and GPU/TPUs, excel at efficiently handling dense, structured data structures by maximizing the utilization of instruction/data pipelines. However, their efficiency diminishes when dealing with sparse, asynchronous event streams~\cite{dave2021hardware}. Lately, a new wave of neuromorphic processors~\cite{amir2017low,davies2018loihi,furber2014spinnaker,hoppner2021spinnaker, khaddam2022hermes, le202364, pei2019towards} has emerged, designed to exhibit low-latency and low-power consumption, suited to model Spiking Neural Networks (\textit{SNNs}) and handle sparse data stream from event-based sensors. Neuromorphic chips are characterized as interconnected many-core systems capable of instantiating vast populations of spiking neurons through highly parallelized and energy-efficient hardware architectures. They have the capability to perform inference based on pre-trained neural networks and process data in a brain-inspired manner - marked by energy-efficient operations conducted in parallel, and asynchronous communication. Another source of efficiency on these platforms stem from the proximity of synapses to where their weights are calculated and updated, reducing data movement. Neuromorphic chips support neurons that generate binary output values (spikes) and enable fast evaluation of the MVM (Matrix Vector Multiplication) operations thorough the use of crossbar arrays and memristors~\cite{rao2023thousands, strukov2008missing} which allow reliable long-term storage of multi-bit quantities as the conductance values and dot-product operations as an analog primitive via Kirchoff's law. MVM operations are fundamental to deep learning algorithms, playing a crucial role in their efficient training and deployment. 

Most recent strategies for training SNNs prominently leverage gradient calculations and error backpropagation~\cite{bohte2000spikeprop, eshraghian2021training, 9597475, li2021differentiable, rathi2023exploring, werbos1990backpropagation,wu2018enabling}.
Error backpropagation~\cite{hecht1992theory} is a widely-used training algorithm in deep learning that minimizes a loss function by adjusting the weights of a neural network through successive iterations, allowing complex neural networks to learn hierarchical representations and approximate intricate functions. Implementing error backpropagation for training SNNs on neuromorphic systems is highly resource-intensive, requiring significant computational power and memory usage across multiple time steps per neuron. On the other hand, training deep neural networks offline using GPU/TPU acceleration and converting them to SNNs could overlook the intrinsic neural dynamics and temporal coding inherent to SNNs, resulting in reduced accuracy and increased latency~\cite{cao2015spiking, han2020rmp, lee2016training, rathi2021diet, 7280696, shrestha2018slayer}, necessitating a re-evaluation of SNN training strategies and algorithmic optimizations to fully harness the advantages of these sparse architectures.

Sparse coding is believed to be a key mechanism by which biological neural systems can efficiently process large amounts of complex sensory data~\cite{olshausen1996emergence,olshausen1997sparse,5456194}. It plays a significant rule in human perception. Neurons involved in vision form an over-complete dictionary of signal elements such as texture, orientation, scale, etc., and the spikes emitted in response to a given input stimulus are highly sparse~\cite{olshausen1997sparse}. One implementation of sparse coding is achieved through the use of Locally Competitive Algorithms (LCA)~\cite{rozell2008sparse}. LCA is a computational model and learning algorithm that iteratively updates the activity of neurons to find a sparse representation of the input data. The competitive mechanism within LCA ensures that only a limited number of neurons become active at any given time, enabling sparsity and efficient coding of high-dimensional data. While LCA has been studied as a neuromorphic-implementation-friendly\cite{davies2018loihi,amir2017low,merolla2014million,sheridan2017sparse,teti2022lcanets,parpart2023implementing} approach for sparse representations, it has not been studied as a direct SNN classifier.

\begin{figure}
\centering
\includegraphics[scale=0.5]{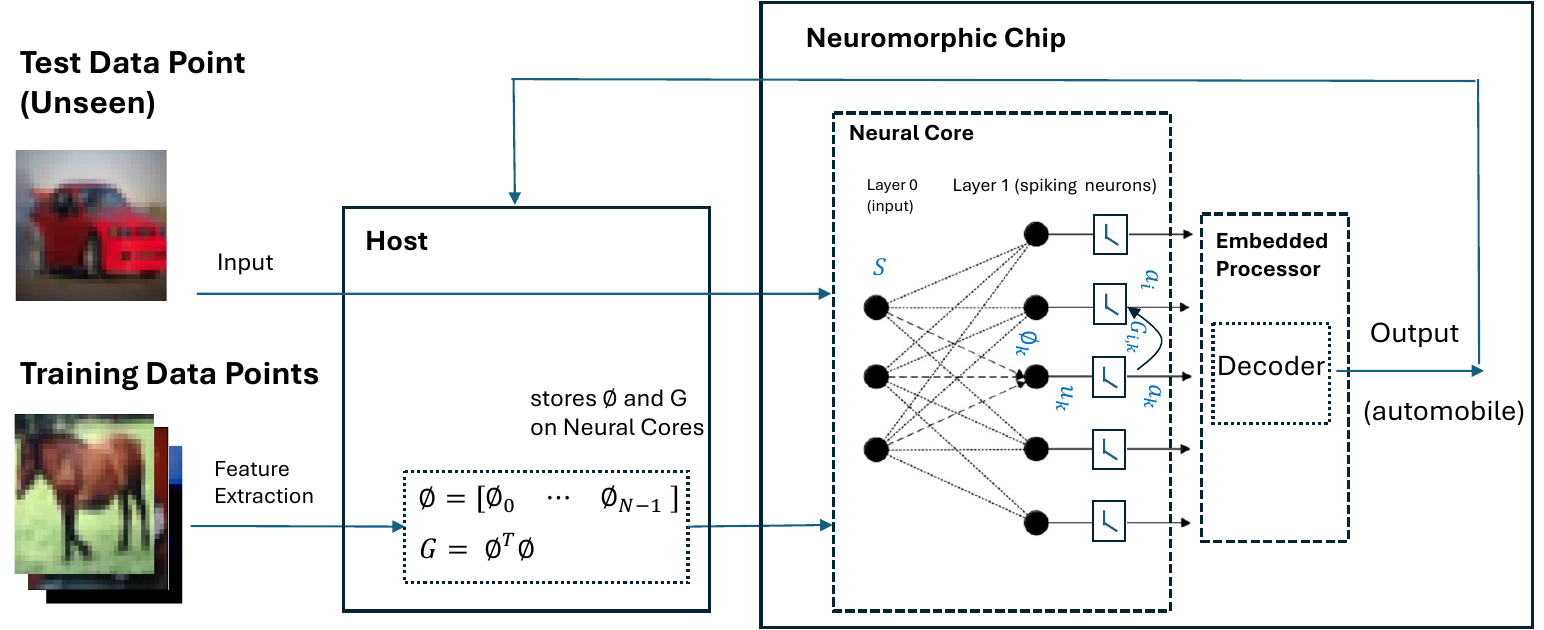}
\caption{Dataset-Scalable Exemplar LCA-Decoder architecture. The features ($\phi_k$) are extracted once and stored on neuromorphic platforms. The spiking neurons have membrane potential $u_k$ and activation $a_k$, and \textit{S} is the input vector.}
\label{lca-arch}
\end{figure}

In this paper, we present the ``Dataset-Scalable Exemplar LCA-Decoder" (\textit{\textbf{D-SELD}}), a novel approach to LCA that approximates the input signal by expressing it as a linear combination of dictionary atoms representing features learned from a training dataset. We then utilize this sparsely encoded set of codes in the ``Decoder" for classification tasks (see Fig.~\ref{lca-arch}). The features are extracted once and stored on neuromorphic platforms, which prioritize specialized operations and energy-efficient computation, contrasting with GPUs or TPUs that focus on high-throughput tasks. 

D-SELD is derived from fundamental principles, has high convergence rate and reduces the need for extensive gradient computations and memory utilization during training. Additionally, the sparse representation of stimuli minimizes the number of firing neurons needed to represent and encode a given input. This introduces sparsity to the network and reduces the computational and memory demands typically associated with training SNNs. 

The D-SELD framework we present showcases promising classification results, excelling in both input reconstruction and the accuracy of classification tasks across well-established benchmark datasets. 
In summary, the contributions of this paper include:
\begin{itemize}
    \item We introduce a dataset-size scalable training methodology for neuromorphic platforms, which mitigates the computational and memory demands inherent in training SNNs using error backpropagation.
    \item We achieved the highest reported top-1 accuracy of 80.75\% on the ImageNet ILSVRC2012 validation dataset and 79.32\% on CIFAR-100 using SNNs.
    \item Our training framework can accommodate the introduction of new training examples in time proportional to the number of new examples, not to the new resulting dataset size, easing post-deployment or task-adjustment of the network for datasets of any size.
\end{itemize}

\section{Related Work}
\label{sec:realted}

As the field of neuromorphic computing continues to advance, novel training methods are being investigated to harness the potential of SNNs in various neuromorphic platforms~\cite{amir2017low,davies2018loihi,furber2014spinnaker,hoppner2021spinnaker, khaddam2022hermes, le202364, pei2019towards}. Here, we will review several key approaches used in training SNNs that are scalable and applicable to larger datasets, followed by an exploration of works on LCA and its deployment on neuromorphic platforms.

\vspace{4pt}\noindent$\bullet$\textbf{~Transfer Learning and ANN-to-SNN Conversion:} Transfer learning is a technique in machine learning that leverages knowledge from one domain to assist in learning another but related task~\cite{bello2021revisiting, donahue2014decaf, girshick2014rich, kornblith2019better, sharif2014cnn, sharma2018analysis, torrey2010transfer, weiss2016survey}. Initially, a model is trained on a large dataset such as ImageNet~\cite{deng2009imagenet} to extract general features and patterns relevant to its primary task. Subsequently, the pretrained model undergoes fine-tuning on a smaller dataset specific to the new task. For example, Sharma, et al.~\cite{sharma2018analysis} reported that ResNet-50 achieved top-1 accuracy rates of 78.10\% on CIFAR-10 and 59.82\% on CIFAR-100. More recently, Bello et al.~\cite{bello2021revisiting} achieved a top-1 accuracy of 88.1\% by fine-tuning a ResNet-152 model on the CIFAR-100 dataset. 
Transfer learning has also been applied to on-chip learning on edge devices with stringent energy and memory limitations~\cite{cai2020tinytl, whatmough2019fixynn, zhu2023pockengine}. For instance, \cite{whatmough2019fixynn} achieved an accuracy of 93.3\% on CIFAR-10~\cite{cifar} and 81.2\% on CIFAR-100 using MobileNet~\cite{mobilenet}.
Zhu et al.~\cite{zhu2023pockengine} employed sparse backpropagation to fine-tune a ResNet-50 model on the CIFAR-10 dataset, achieving an accuracy of 96.2\%. 

In the context of SNNs, transfer learning typically involves training an ANN with ReLU neurons using supervised learning algorithms. Afterwards, an SNN with spiking neurons and a similar architecture to the source ANN model is created, initializing the SNN with weights learned from the pretrained ANN model~\cite{cao2015spiking,han2020rmp, rathi2021diet, 7280696, yan2021near}. This approach does not explicitly consider neuron dynamics during training, leading to reduced accuracy and increased latency compared to the original ANN. Moreover, fine-tuning requires a significant memory footprint during error backpropagation, making it impractical for the training computation to be performed on neuromorphic devices.

Incorporating learned feature maps into D-SELD introduces a novel approach to transfer learning. This method utilizes features extracted from the penultimate layer of  ubiquitous CNN models to construct a dictionary for LCA encoding, enabling a straightforward decoder to operate effectively as a classifier. This innovation paves the way for deploying larger models on neuromorphic platforms.

\vspace{4pt}\noindent$\bullet$\textbf{~Direct SNN Training:} Several methods perform credit assignment in SNNs using error backpropagation~\cite{bohte2000spikeprop, esser2015backpropagation, guo2022loss, lee2016training,li2021differentiable,perez2021sparse,shrestha2018slayer,xiao2022online,superspike,zenke2021remarkable}. Directly implementing backpropagation in SNNs is however challenging due to the discontinuous and non-differentiable nature of spiking neurons. To address this issue, surrogate gradients~\cite{lee2016training,8891809,zenke2021remarkable, neftci2019surrogate} are used to approximate the discontinuous derivative as a continuous function, enabling end-to-end backpropagation training of SNNs. The utilization of error backpropagation still demands massive computations and memory, making the training process resource-intensive. Despite its success in training SNNs, the implementation of error backpropagation for on-chip training has been delayed due to its substantial demands on computational and memory resources.

D-SELD eliminates the need for any gradient computations associated with training. Additionally, the sparse encoding of stimuli with LCA, minimizes the number of firing neurons needed to represent a given input. This introduces sparsity to the network which further reduces the computational demands.

\vspace{4pt}\noindent$\bullet$\textbf{Sparse coding and LCA:}
Sparse coding and applications of LCA have been extensively studied in the past~\cite{lee2006efficient, mairal2014sparse, rigamonti2011sparse, teti2022lcanets, zhang2015survey}.
Several approaches such as pruning~\cite{abs-1803-03635,He_2018_ECCV,23864}, regularization~\cite{girosi1995regularization,wright2008robust}, dropout~\cite{srivastava2014dropout}, and probabilistic quantization~\cite{guo2018survey} have been employed in ANNs to promote sparsity. While LCA shares similarities with regularization techniques through its utilization of $\ell_1$-minimization, its distinct incorporation of spiking behavior sets it apart, making it well-suited for neuromorphic deployment. Recent implementations of sparse coding and LCA on neuromorphic platforms such as Loihi~\cite{davies2018loihi} and TrueNorth~\cite{amir2017low, merolla2014million} and others~\cite{sheridan2017sparse} serve to further validate this advantage. 

\vspace{4pt}\noindent$\bullet$\textbf{Neuromorphic Deployment:}
The feasibility of employing the original LCA on the IBM TrueNorth Neuromorphic platform has been explored in~\cite{fair2019sparse}. The experimental findings indicate that the LCA can be energy-efficiently implemented on this neuromorphic computing platform to perform sparse coding. The results demonstrate comparable performance to conventional CPUs, suggesting the potential of LCA for energy-efficient sparse representation on neuromorphic platforms. The LCA has also been deployed on the Loihi platforms to solve LASSO optimization problems with over three orders of magnitude superior energy-delay product compared to conventional solvers running on a CPU~\cite{davies2018loihi}. 
Furthermore,~\cite{frady2020neuromorphic} demonstrated K-Nearest Neighbor search using sparse representation of images on a Pohoiki Springs system.
In addition, reference~\cite{parpart2023implementing} provides a comprehensive benchmarking of LCA on Loihi2.

\section{Dataset-Scalable Exemplar LCA-Decoder Design}\vspace{-8pt}
\label{sec:design}
In this section, we begin by providing an overview of sparse coding and the LCA algorithm. Then, we introduce the Dataset-Scalable Exemplar LCA-Decoder framework and derive the equations that govern its operations.

\subsection{Sparse coding and LCA}
In the sparse coding algorithm~\cite{rozell2008sparse}, the input signal $\mathit{S}$ is represented as:
\begin{equation}
\label{recon}
S = \phi a + \varepsilon 
= \sum_{i=0}^{M-1} \phi_{i} a_{i}  + \varepsilon
\end{equation}
where $\phi$ is a dictionary of features $\phi_i$ and $a$ is a vector of activation coefficients $a_i$. The term $\varepsilon$ represents Gaussian noise. The dictionary $\phi$ is over-complete, meaning that the number of columns in $\phi$ (\textit{M}) is greater than the number of rows \textit{N}, and as a result, it is non-orthogonal. This non-orthogonality leads to an infinite number of possible combinations, and each combination is an equally valid representation of signal $\mathit{S}$. Prior work has demonstrated that increasing over-completeness leads to higher spatial frequency features and a more even distribution of orientations on natural images~\cite{paiton2019analysis}.

One implementation of sparse coding that minimizes the energy function (Eq.~\ref{energy function}) is through the Locally Competitive Algorithm (LCA). In LCA, activation coefficients $a_{i}$ correspond to the outputs of Leaky Integrator and Fire (LIF) neurons, defined by equations \ref{neuron update} to \ref{thresholding}. The primary objective of LCA is to minimize reconstruction error while maintaining an acceptable level of network sparsity. The reconstructed input $\hat{S}$ is expressed as $\hat{S}=\sum_{i=0}^{M-1}\phi_{i}a_{i}$, where $\phi_i$ are basis functions or dictionary atoms and, $a_{i}$ represents the activation of the LIF neuron.

\begin{equation}
\label{energy function}
E = \underset{Reconstruction}{\underbrace{\frac{1}{2}\begin{Vmatrix}
S-\widehat{S}
\end{Vmatrix}_{2}^{2}}} + \underset{Sparsity}{\underbrace{\lambda \sum_{i=0}^{M-1}\left | a_{i} \right |}}
\end{equation}

The LIF neuron's membrane potential $u_i$ is subject to a driving excitatory input $b_i$ and an inhibition matrix (Gramian) $G$:
\begin{equation}
\label{neuron update}
\tau \dot{u}_i[k]+u_i[k]=b_i-\sum_{m\neq i}^{M-1}G_{i,m}a_m[k]
\end{equation}

\begin{equation}
b_i = S\phi_i
\end{equation}

\begin{equation}
G = \phi^{T}\phi
\end{equation}
The Gramian matrix allows stronger neurons to prevent weaker neurons from becoming active, resulting in a sparse representation. Specifically, the inhibition signal from the active neuron \textit{m} to any other neuron \textit{i} is proportional to the activity level $a_m$ of neuron \textit{m} and to the inner product between the two neurons' receptive fields (proportional dictionary elements). And, finally, the thresholding function is:
\begin{align}
\label{thresholding}
    a_i[k] &= T_\lambda(u_i[k]) \nonumber \\
    &=
    \begin{cases}
        u_i[k] - \lambda \operatorname{sign}(u_i[k]), & \quad \left| u_i[k] \right| \geq \lambda \\
        0, & \quad \left| u_i[k] \right| < \lambda
    \end{cases}
\end{align}
Here, the threshold `$\lambda$' refers to the level that the membrane potential must exceed for the neuron to become active. This threshold is the same sparsity penalty trade-off utilized in the energy function (Eq.~\ref{energy function}).
After obtaining the neuron activations for an initial dictionary $\phi$, these activations are utilized to update the dictionary using Eq.~\ref{dict_update} where $\eta$ is the dictionary learning rate:
\begin{align}
\label{dict_update}
\phi_i = \phi_i + \eta (S -\hat{S})a_i
\end{align}


\subsection{Dataset-Scalable Exemplar LCA-Decoder Framework}
\label{Exemplar-LCA-Decoder}

To leverage the efficiency of LCA and the sparsity of spiking neurons for future deployment on neuromorphic hardware platforms, we adopt a novel strategy: constructing the dictionary $\phi$ directly from a collection of features extracted from the training dataset, thus bypassing the costly process of dictionary learning. In other words, each $\phi_i$ will represent a feature learned from a data point $x_i$ in the training dataset. This can be expressed as:
\begin{align}
\phi = [\phi_0, \phi_1, \ldots, \phi_{M-1}]
\end{align}
where $\phi_i = f(x_i)$, and $f(\cdot)$ denotes the function that extracts features from $x_i$. This eliminates the need for image reconstruction $\hat{S}$ and dictionary updates in Eq.~\ref{dict_update}, differing from original LCA, where Stochastic Gradient Descent (SGD) could be used to learn and update the dictionary for each batch of input data. 

Next, we'll map any unseen test data $S_{test}$ to the resulting \textit{M}-dimensional space of feature vectors (dictionary atoms $\phi_i$) and find activation codes $a_i$ that approximate the new input using Eq.~\ref{thresholding}.

\subsubsection{Feature Extraction}
\label{feature extraction}
While constructing the dictionary $\phi$ directly from the training dataset suffices for image reconstruction tasks and classification of simpler datasets like MNIST~\cite{lecun1998mnist}, achieving higher classification accuracy on more complex datasets requires the use of more complex features from the training dataset. 
Prior works have demonstrated that features extracted from convolutional layers in Convolutional Neural Networks (CNNs) exhibit rich characteristics~\cite{donahue2014decaf, girshick2014rich, sharif2014cnn}. Additionally, while these models are typically trained on large-scale image datasets such as ImageNet~\cite{deng2009imagenet}, the extracted features have proven effective across various other datasets. We investigated several off-the-shelf CNN models such as MobileNet~\cite{mobilenet}, DenseNet~\cite{densenet}, ResNet~\cite{resnet} and EfficientNet~\cite{tan2021efficientnetv2}, renowned for their demonstrated efficacy on ImageNet dataset. 

\begin{figure}
  \centering
  \begin{subfigure}{0.45\textwidth}
    \centering
    \includegraphics[width=0.8\textwidth]{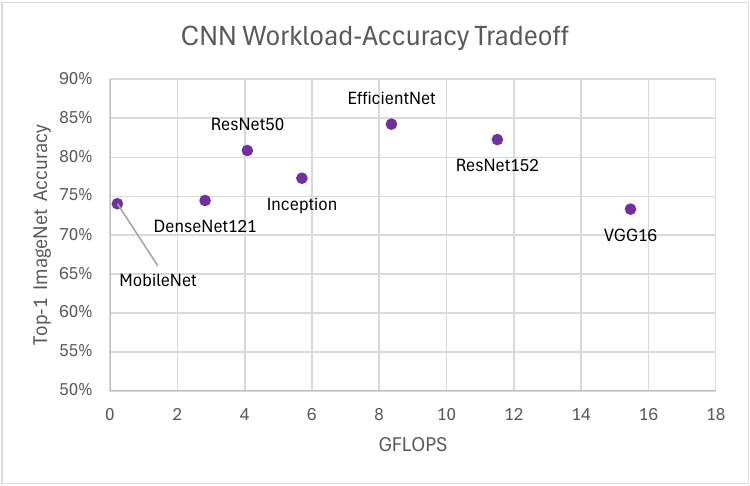}
    \caption{The numerical values are sourced from reference~\cite{torchvision}}
    \label{CNN models}
  \end{subfigure}
  \hfill
  \begin{subfigure}{0.45\textwidth}
    \centering
    \includegraphics[width=\textwidth]{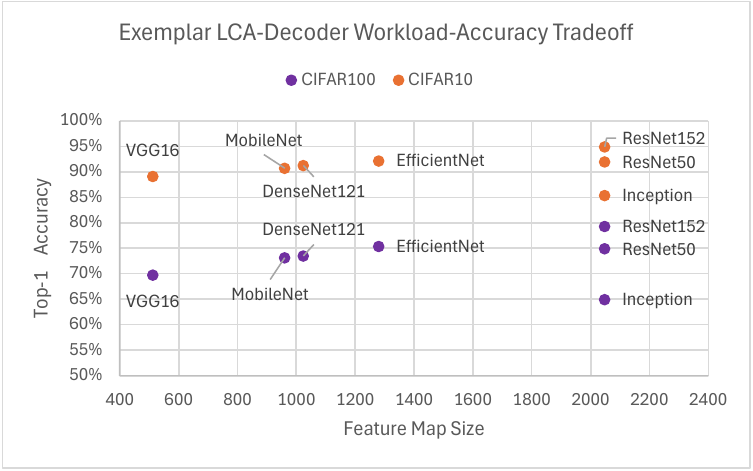}
    \caption{}
    \label{FM-Sizes}
  \end{subfigure}
  \hfill
  \caption{Trade-off between accuracy and workload across CNN models used as feature extractors in D-SELD}
\end{figure}

Figure~\ref{CNN models} shows the trade-off between top-1 accuracy on ImageNet dataset and the reported computational workload across these models~\cite{torchvision}. GFLOPS are estimates of the number of (Giga) floating point operations performed during a single full pass of a model. Among these models, EfficientNet achieves the highest accuracy, while MobileNet has the lowest number of GFLOPS. Despite variations in architecture, accuracy, and workload, the features extracted from these models can be uniformly utilized in D-SELD. The primary distinction lies in the sizes of their feature maps. Since the features are extracted once prior to deployment on the neuromorphic hardware, the size of feature maps provides a more accurate indication of the workload for D-SELD. Figure~\ref{FM-Sizes} illustrates the trade-off between workload and accuracy for D-SELD, using feature map sizes as a metric. Features extracted from the ResNet-152 architecture, with a feature size of 2048, achieved the highest accuracy on CIFAR-10~\cite{cifar} and CIFAR-100 datasets, followed by features from EfficientNet with a feature size of 1280. With a feature map size of 512, VGG-16 achieved the lowest accuracy. Since all models utilized the same decoder, the superior performance of ResNet-152 in transfer learning could be attributed to its ability to extract more discriminative features.

\subsubsection{Design Considerations for Decoding Algorithms}
\label{decoding}
In this section, we develop a decoder aimed at decoding sparse codes $a_i$ generated by Exemplar LCA and map them to \textit{K} distinct object classes from the training dataset. Later, we'll employ the same decoder to accurately predict the class of unseen test data points. 

With an adequate number of training data points \textit{M}, any data point will have a sparse coding representation as $a = [a_0, a_1, a_2, ..., a_{M-1}]$. In an ideal scenario, for any input \textit{S}, the non-zero entries in vector \textit{a} will exclusively correspond to the dictionary atoms $\phi_i$ belonging to a singular object class \textit{k} where \textit{k} ranges from 1 to \textit{K}. However, the presence of modeling errors and noise may result in small non-zero entries that are associated with multiple object classes. We will design decoders that can resolve this issue:

\paragraph{I. Maximum Activation Code}
For a given unseen test input $S_{Test} = \sum_{i=0}^{M-1} \phi_{i}a_{i} + \epsilon$, we calculate sparse activation codes $a_i$ and determine the class $C_{Test}$ as the class associated with the maximum value in the sparse coding set of $S_{Test}$:
\begin{equation}
C_{Test} = Class[argmax_{i} [a_{i}]],
\end{equation}

\paragraph{II. Maximum Sum of Activation Codes}
To better harness the relevance of neuron activations within each class, we sum the $\ell_1$ norms of the activations for each object class \textit{k}, and assign $S_{Test}= \sum_{i=0}^{M-1} \phi_{i}a_{i} + \epsilon$ to the class with the highest value:
\begin{equation}
\label{max-sum-decoder}
C_{Test} = Class[argmax_{k} \sum \left |  a_{i}^{(k)}\right |] 
\end{equation}

This approach yields significantly improved results compared to the Maximum Activation decoder. However, to fully leverage the discriminative coding among various object classes, we also explored the application of a shallow neural network.

\paragraph{III. Shallow Neural Network}
Here, we employ a shallow network (single-layer perceptron) to map sparse activation codes $a_i$ to \textit{K} class labels. Sparse activation codes, which represent the presence or absence of specific features or patterns, serve as inputs to the network. Through the process of training, the network learns to associate these sparse activations with corresponding class labels. By adjusting the weights within the network, it can discern patterns in the sparse activation space and more accurately predict the class labels of unseen test data. This approach incurs additional memory and computational costs compared to the previous two encoders, and should only be used if previous decoders fail. When extracted features capture rich content from the training dataset, the performance of Maximum Sum of Activation Codes closely approaches that of a Shallow NN decoder (less than a 1\% difference). Therefore, we include the results of this decoder in Table~\ref{tab:decoders} for review and omit them from the benchmarking table and efficiency discussions.

\section{Experimental Results}
\label{sec:evaluation}
In this section, we present the results obtained by applying our LCA-Decoder approach to image classification tasks on MNIST~\cite{lecun1998mnist}, CIFAR-10~\cite{cifar}, CIFAR-100~\cite{cifar} and ImageNet~\cite{deng2009imagenet} datasets. MNIST consists of a training dataset of 60K images and a test dataset of 10K images, each with images sized 28x28 pixels in grayscale. The CIFAR-10 and CIFAR-100 datasets consist of 50K training images and 10K test images, each with dimensions of 32x32 pixels in RGB format. The ImageNet dataset includes 1.28 million training images and 50K validation images. We randomly split the training dataset and selected 50K samples to construct the dictionary. We exclusively used the validation dataset for testing purposes, ensuring it was not used prior to testing. All datasets were obtained using Torchvision~\cite{torchvision}.

\begin{figure}[t]
    \centering
    \includegraphics[width=0.9\textwidth]{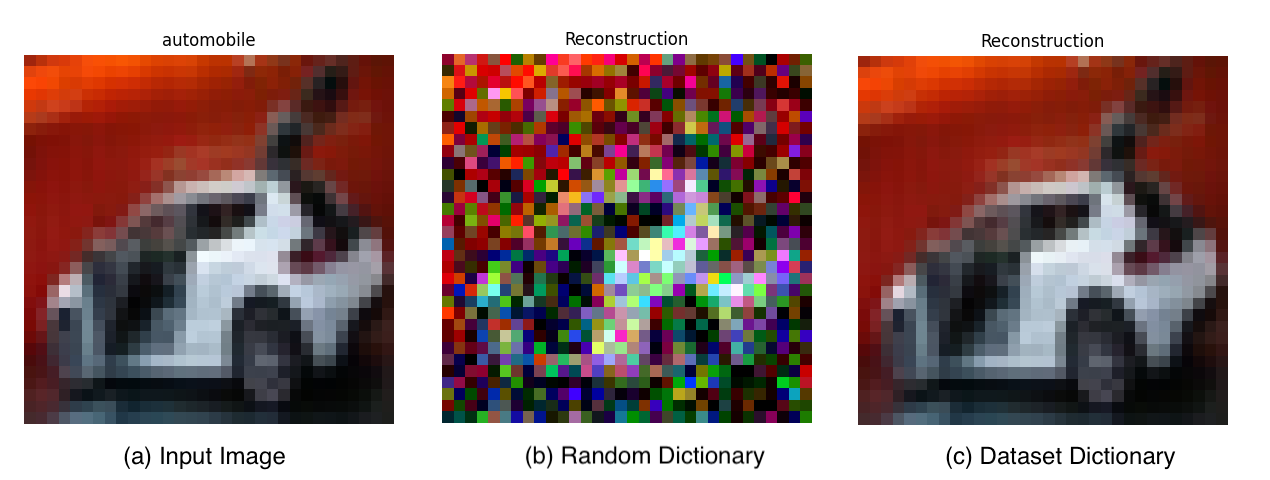}
    \caption{The reconstructed images of an automobile from the CIFAR-10 dataset. The ``Dataset" dictionary performs better in reconstructing the input image. Input image is sourced from unseen test dataset.}
  \label{DictResults}
\end{figure}
\begin{table}[t]
  \centering
  \caption{Reconstructed Image Quality Metrics}
\begin{tabular}{|c|c|c|}
  \hline
   & \textbf{Random Dictionary} & \textbf{Dataset Dictionary} \\
  \hline
  Mean Squared Error (MSE) & 675.3 & 92.45  \\
  \hline
  Peak Signal-to-Noise Ratio (PSNR) & 19.84 dB & 28.47 dB \\
  \hline
  Structural Similarity Index (SSIM) & 0.75 & 0.99  \\
  \hline
    \hline
  Number of Iterations & 40 & 6\\  
  \hline
\end{tabular}
\label{Image Quality Metrics}
\end{table}

\subsection{Image Reconstruction}
\label{subsec:dictionary}
\begin{figure}
  \centering
  \begin{subfigure}{0.45\textwidth}
    \centering
    \includegraphics[width=\textwidth]{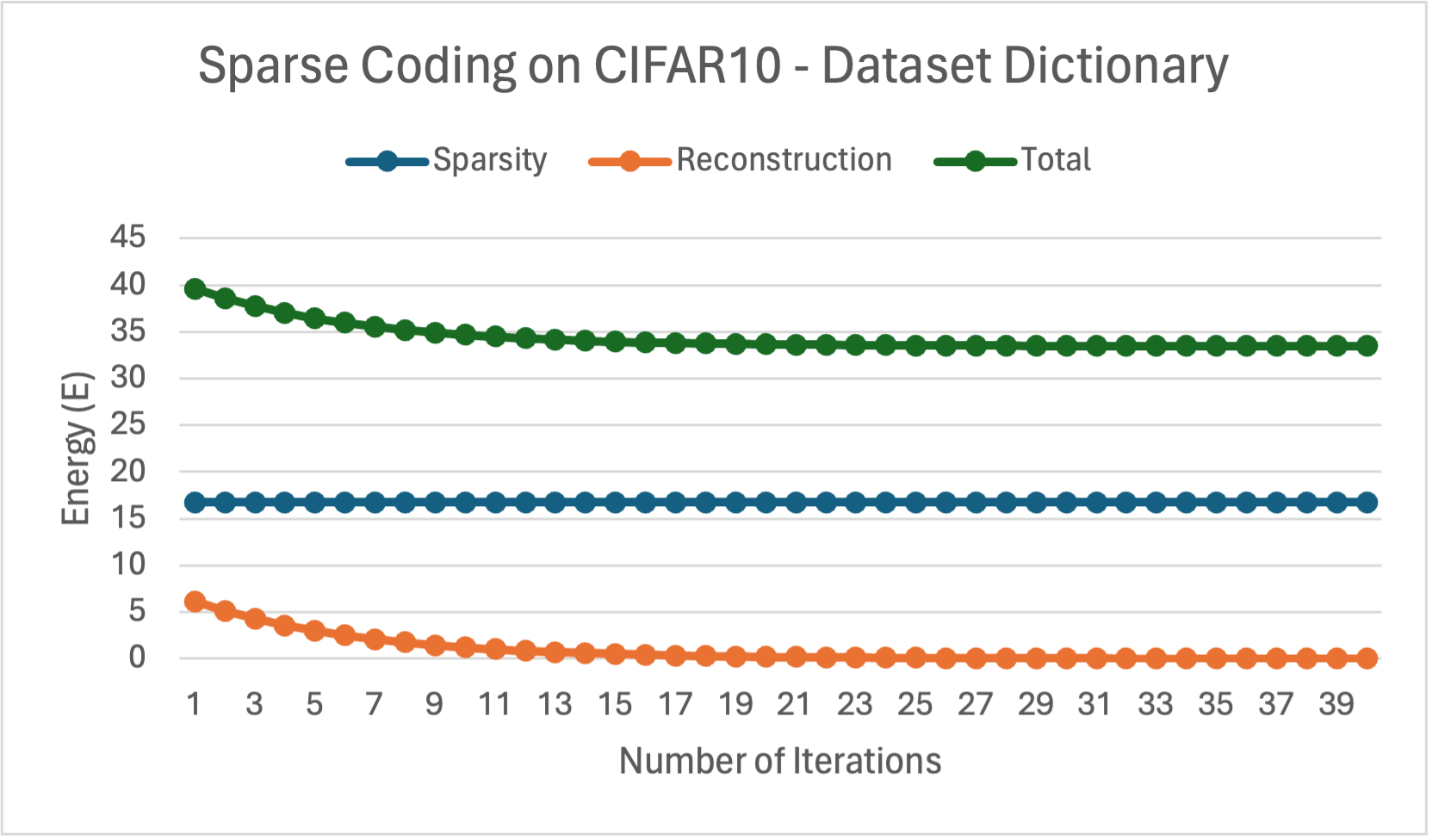}
    \caption{Dataset Dictionary}
    \label{CIFAR10-Dataset:a}
  \end{subfigure}
  \hfill
  \begin{subfigure}{0.45\textwidth}
    \centering
    \includegraphics[width=\textwidth]{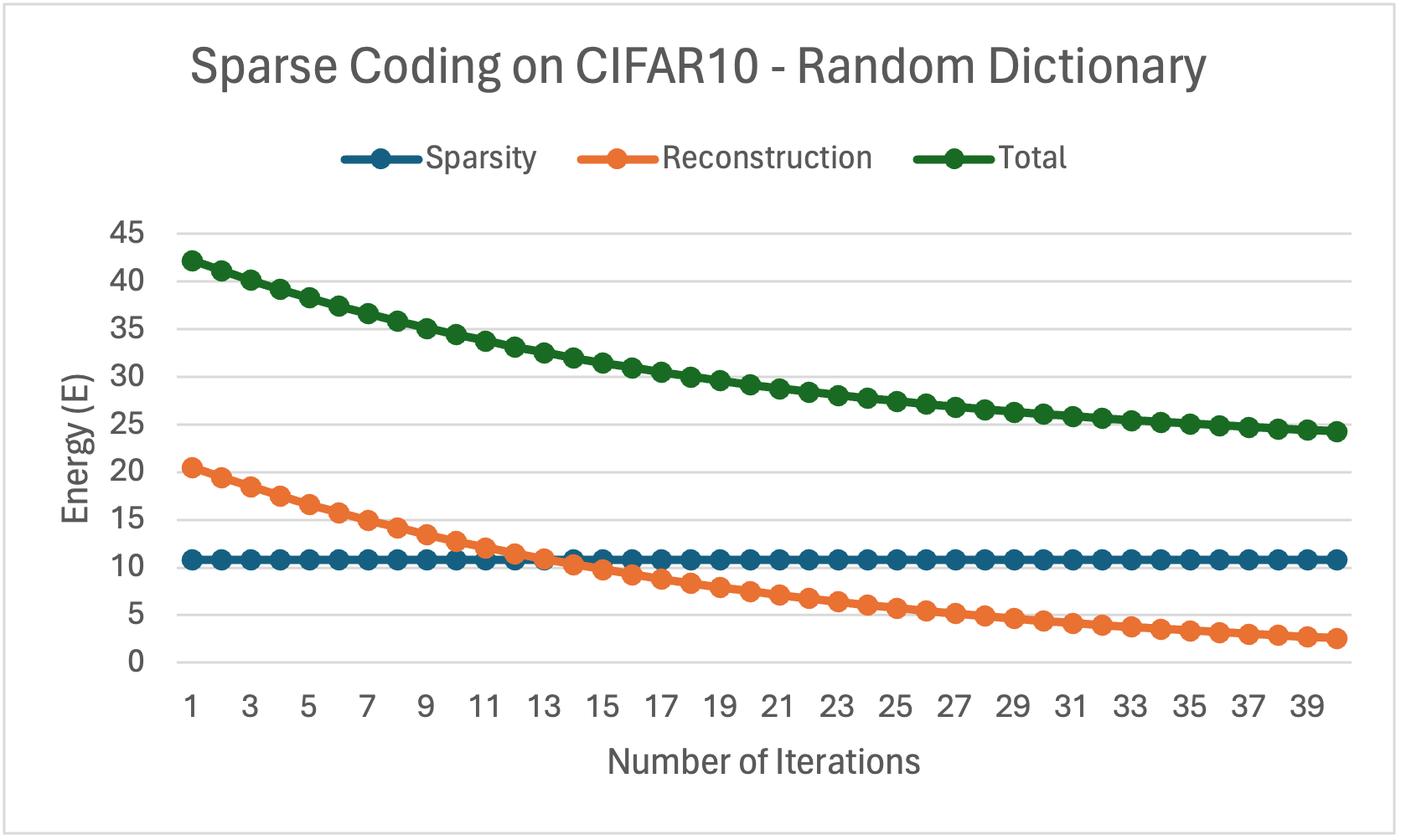}
    \caption{Random Dictionary}
    \label{CIFAR10-Random:b}
  \end{subfigure}
  \hfill
  \caption{Energy function $E$ during image reconstructions (a and b).}
  \label{CIFAR10-SC}
\end{figure}

We initially conducted experiments to assess the effectiveness of exemplar encoding in image reconstruction as a proof of concept. We used unseen images from the MNIST and CIFAR-10 datasets. 
First, we utilized a random dictionary, where the dictionary atoms $\phi_i$ were initialized using random values drawn from a Gaussian distribution N(0, 1). This choice aligns with the common practice of initializing weights to random values in ANNs. 
Additionally, as discussed in Section~\ref{Exemplar-LCA-Decoder}, in the ``Dataset Dictionary", we directly selected dictionary atoms from the training dataset.
The results depicted in Fig.~\ref{DictResults} demonstrate that using the ``Dataset" dictionary surpasses the performance of a ``Random" dictionary by achieving higher reconstruction quality and faster convergence (see Table~\ref{Image Quality Metrics}). 
Fig.~\ref{CIFAR10-SC} illustrates all components of the energy function $E$ during image reconstructions and confirms the convergence of LCA in both cases.  

\begin{figure}
\centering
\includegraphics[scale=0.6]{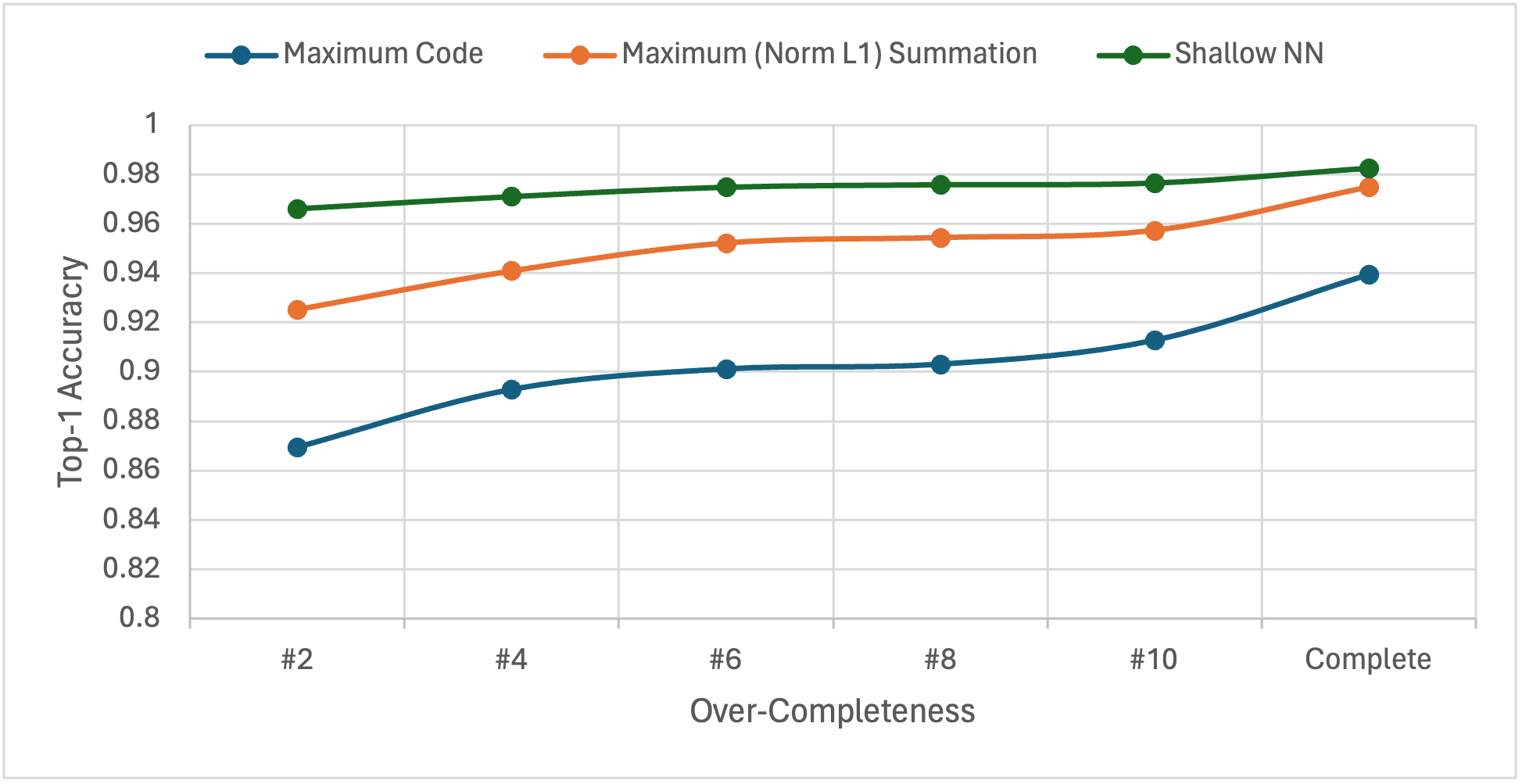}
\caption{Top-1 Accuracy comparison across three decoders with increasing dictionary over-completeness on MNIST dataset (image size: 28 * 28). Here, \#2 represents dictionary size of 28 * 28 * 2 and ``Complete" denotes the utilization of all training data points.}
\label{over-completeness}
\end{figure}

\subsection{Dictionary Size and Over-completeness}
\label{subsec:over-complete} 
 
A dictionary $\phi$ is considered ``over-complete" when the number of dictionary atoms (neurons) is greater than the number of input dimensions (image pixels). Fig.~\ref{over-completeness} shows that in classification tasks, increasing over-completeness is associated with enhanced top-1 accuracy, up to an upper bound defined by the training dataset size. As the size of the dictionary can be adjusted based on the size of the training dataset, this scalability enables the an LCA-Decoder approach to uniformly apply to datasets of any size. Moreover, new training data points or classes can be incrementally added to the dictionary online.

\subsection{Classification Results}

\begin{figure}
  \centering
  \begin{subfigure}{0.475\textwidth}
    \centering
    \includegraphics[width=\textwidth]{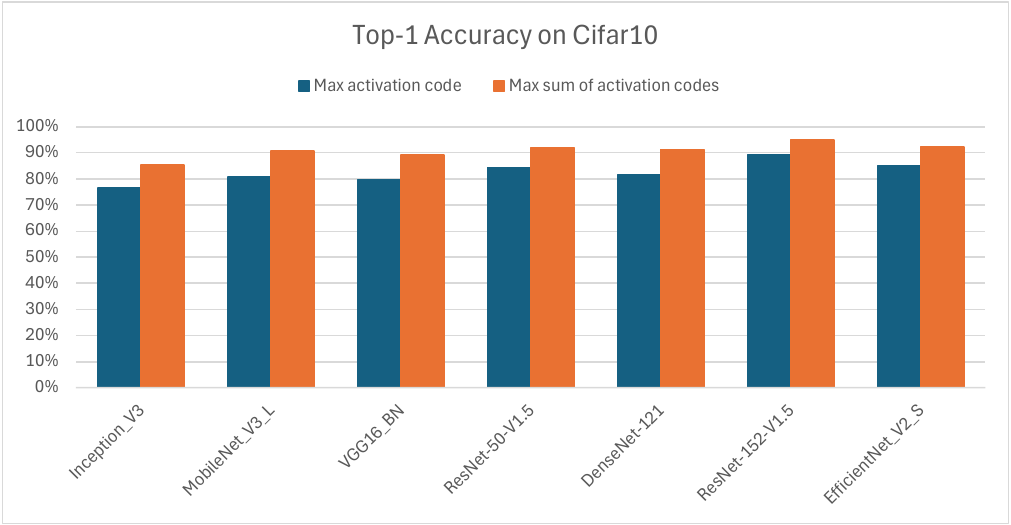}
    \caption{CIFAR-10}
    \label{CIFAR10-All}
  \end{subfigure}
  \hfill
  \begin{subfigure}{0.475\textwidth}
    \centering
    \includegraphics[width=\textwidth]{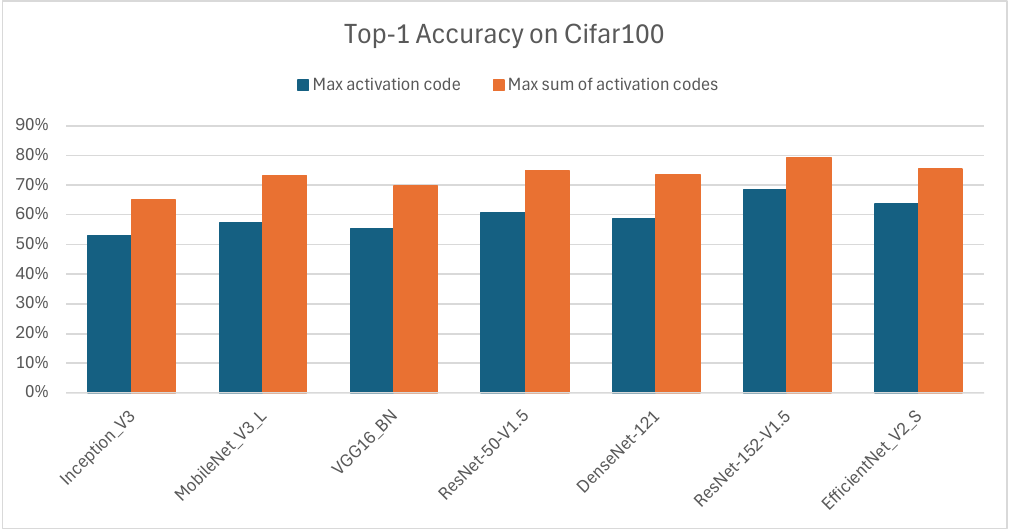}
    \caption{CIFAR-100}
    \label{CIFAR100-All}
  \end{subfigure}
    \hfill
  \begin{subfigure}{0.30\textwidth}
    \centering
    \includegraphics[width=\textwidth]{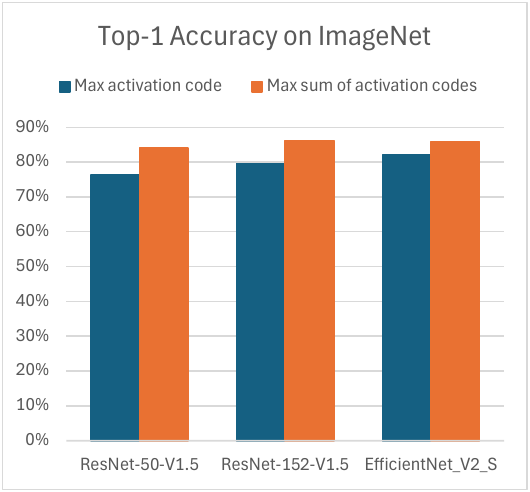}
    \caption{ImageNet}
    \label{ImageNet-All}
  \end{subfigure}
  \hfill
  \caption{Comparison of top-1 accuracy across three datasets using Maximum Activation Code (blue) and Maximum Sum of Activation Codes (orange), with features extracted from various CNN models.}
  \label{fig:all-models}
\end{figure}

\begin{table}
\caption{Hyperparameters}
\label{parameter}
\begin{center}\small
\begin{tabular}{|c|c|c|}
\hline
\textbf{Symbols} & \textbf{Description} & \textbf{Value} \\
\hline
$\lambda$ & Threshold & 2 \\
$M$ & Dictionary Size & 50K \\
$\tau$ & Leakage &  100\\
$k$ & Number of time steps &  100\\
\hline
\end{tabular}
\end{center}
\end{table}

\begin{table}
  \centering
  \caption{Comparison of Top-1 Accuracy Scores.}
    \begin{tabular}{|l|>{\centering\arraybackslash}m{2.75cm}|>{\centering\arraybackslash}m{2cm}|c|>{\centering\arraybackslash}m{2.25cm}|>{\centering\arraybackslash}m{2.25cm}|}
    \cline{4-6}
    \multicolumn{2}{c}{} & \multicolumn{1}{c}{} & \multicolumn{3}{|c|}{\textbf{Decoding Methods}}  \\
    \cline{4-6}
      \multicolumn{3}{c}{} & \multicolumn{2}{|c|}{\textit{D-SELD}} & \multicolumn{1}{c|}{ \textit{D-SELD+BP} } \\
      \hline
      \textbf {}& \textbf{Feature Extractor} & \textbf{Feature Map Size} & \textbf {$\underset{i}{max} \left\{ a_{i} \right\}$} & \textbf {$\underset{k}{max} \sum_{i} \left |  a_{i}^{(k)}\right |$} & Shallow NN\\
    \hline
      MNIST& none & 28 * 28 * 2& 93.93\% & 97.48\% & \textbf{98.25\%}\\
      \hline
        CIFAR-10 & ResNet-152-V2& 2048& 89.16\% & 94.98\% & \textbf{95.19\%}\\
      \hline   
     CIFAR-100& ResNet-152-V2& 2048 & 68.29\% & 79.32\% & \textbf{79.33\%}\\
      \hline 
     ImageNet& ResNet-152-V2& 2048 & 71.83\% & \textbf{80.75\%} & 21.60\%\\
      \hline 
     ImageNet& EfficientNet-V2-S & 1280 & 76.40\% & \textbf{80.68\%} & 21.49\%\\
      \hline     
    \end{tabular}
        \label{tab:decoders}
\end{table}
Figure~\ref{fig:all-models} shows the top-1 accuracy performance. All experiments were consistently conducted using the hyperparameters specified in Table~\ref{parameter}. In all cases, using the Maximum Sum of Activation Codes decoder resulted in higher accuracy compared to using the Maximum Activation Codes decoder. Among the tested CNN feature extractors, ResNet-152-V2~\cite{resnet50v1.5} consistently achieved the highest accuracy. Further hyperparameter optimization and fine-tuning threshold and/or leakage are required to achieve accuracy comparable to or higher than that reported at K=100. For instance, setting a threshold of 0.3 and $\tau$ to 100 achieves 74.26\% accuracy on CIFAR100 using ResNet152.

Table~\ref{tab:decoders} shows the highest top-1 accuracy scores achieved for tested datasets using all decoders described in Section~\ref{decoding}. For MNIST, data points were used directly without applying any feature extractor. For the CIFAR-10, CIFAR-100 and ImageNet datasets, we achieved the highest accuracy using features extracted through ResNet-152-V2 architecture. And among the tested decoders, shallow NN decoder achieved slightly higher accuracy on MNIST, CIFAR-10 and CIFAR-100 datasets, while the maximum sum of sparse codes for each class resulted in the highest accuracy for ImageNet.

\subsection{Sparsity}
\label{sparsity}
\begin{figure}
\centering
\includegraphics[width=0.5\textwidth]{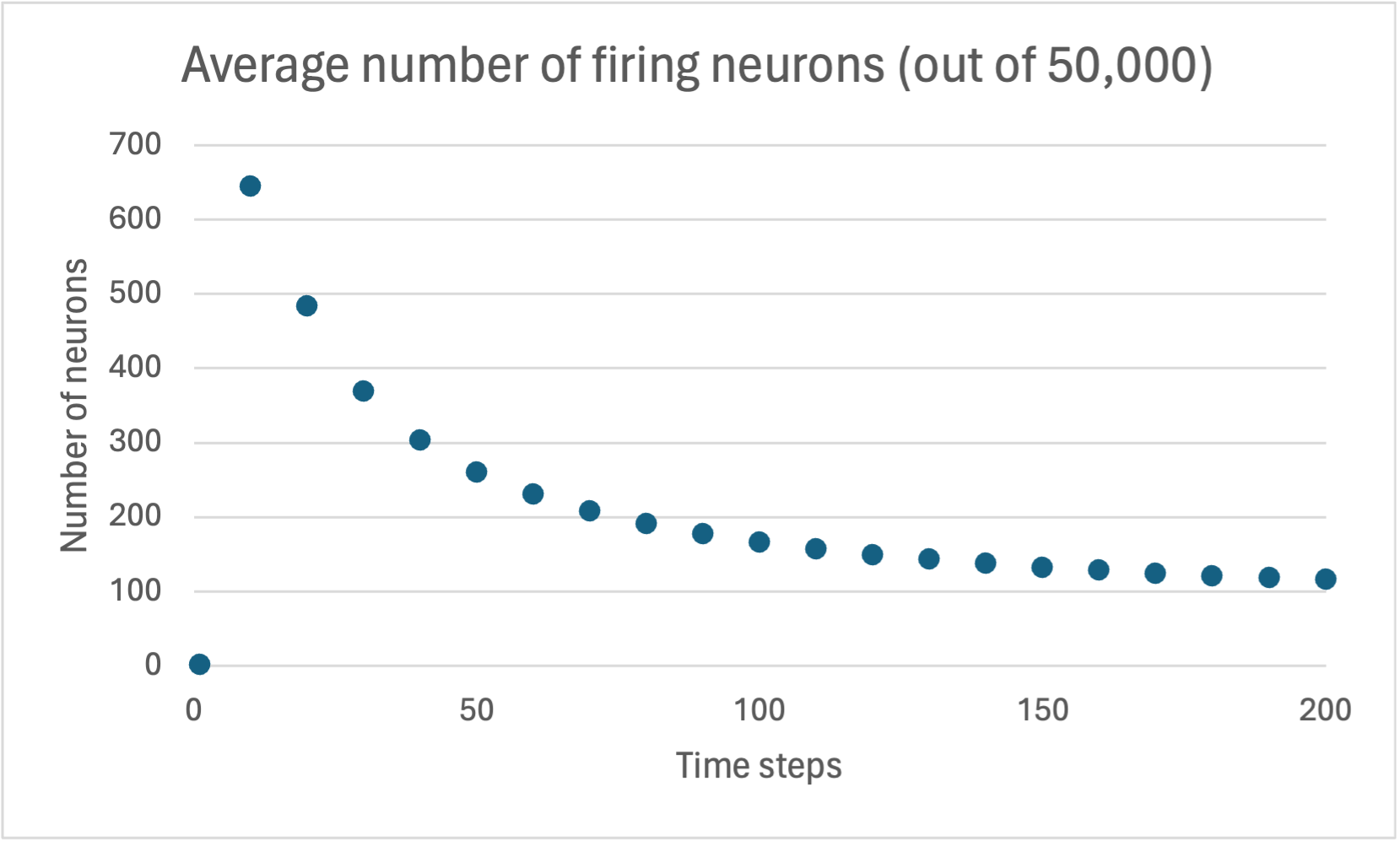}
\caption{Average firing neurons count for Dataset-Scalable Exemplar LCA-Decoder on CIFAR-100}
\label{fig:sparsity}
\end{figure}

Sparsity in D-SELD arises due to firing neurons transmitting inhibition signals to non-firing neurons, impeding their ability to reach the activation threshold. To assess D-SELD's sparsity, we examine the proportion of neurons within the encoding layer that fire during the processing of input data.
Figure~\ref{fig:sparsity} illustrates the average number of firing neurons upon applying the Exemplar LCA-Decoder on CIFAR-100 with 10,000 test samples. We noted that only a minuscule (less than 1\% with an average value of 0.4\%) proportion of the 50,000 neurons were fond to fire at each time step. Since each firing neuron only fires one spike at each time step, this plot also shows the number of spikes occurring in each time step. While this number reaches a maximum at early time steps, it declines exponentially over subsequent time steps.  

\subsection{Workload Efficiency Analysis}
Here, we use FLOPs (Floating-Point Operations) as a metric to analyze computational efficiency. Since D-SELD eliminates the need for dictionary learning, the calculations reduce to Equations~\ref{neuron update},~\ref{thresholding} and ~\ref{max-sum-decoder}. Computing $b_i$ requires $N*M$ multiplications and $(N-1)*M$ addition, where $N$ is the feature map size and $M$ is the length of the dictionary (also the number of neurons). The inhibition signal involves $M^2-M$ multiplications and $M^2 -2M$ additions. The leakage term adds $M$ multiplication operations, and $3M$ additions/subtractions combine these terms and update the neurons' membrane potentials. Additionally, the Gramian matrix $G$ requires $\frac{M(M+1)N}{2}$ multiplications and $\frac{M(M+1)(N-1)}{2}$ additions. This is considered the training cost, computed once per task (dataset), and excluded from the inference cost.
As a result, the total floating point operations with a time step $K$ are $K(\frac{(2N-1)M}{K}+2M^2+M)$. Note that $b_i$ is computed once per input image and remains constant across iterations. Next we factor in the observed sparsity from section~\ref{sparsity} by redefining $M$ associated with firing neurons as $\hat{M}$.

\begin{align}
\label{FLOPS}
FLOPs = K(\frac{(2N-1)M}{K}+2M\hat{M}+M)
\end{align}

Table~\ref{flops} displays the estimated FLOPs for the exemplary LCA-Decoder. According to the findings in Section~\ref{sparsity}, we assumed that, on average, 0.4\% of neurons ($\hat{M}$) spike at each time step for each model. ``Training (TFLOPs)" denotes the TeraFLOPs required to compute the Gramian matrix, while ``Inference (GFLOPs)" represents the GigaFLOPs for inference operations. We conducted two scenarios with 100 and 10 time steps \textit{K}. Reducing the number of time steps from 100 to 10 resulted in an average reduction of 84\% in Inference GFLOPs. VGG-16 has the lowest computation workload among the tested models, as anticipated from Fig.~\ref{FM-Sizes}.

We conducted two experiments on the NVIDIA A100 and Apple MacBook, averaging execution times over 5 runs per experiment. Table~\ref{inference-time} displays the measured execution times for individual components of D-SELD. In this experiment, features were extracted from the ResNet-152 architecture and used for classifying the CIFAR-100 dataset. We executed batches of 64 on the A100 GPU; however, the reported numbers correspond to a single test data. As evident, the computational bottleneck resides in the LCA encoding phase, which stands to gain significant advantages from hardware specialization in neuromorphic chips. 

\begin{table}[ht]
\caption{Estimated workload analysis of D-SELD}
\label{flops}
\centering
\begin{tabular}{|l|>{\centering\arraybackslash}m{2.5cm}|l|l|l|l|>{\centering\arraybackslash}m{2.5cm}|}
\hline
 \textbf{Feature Extractor} & \textbf{Training (TFLOPs)} &\textbf{K} & \textbf{N} & \textbf{M} & $\bm{\hat{M}}$ & \textbf{Inference (GFLOPs)}\\ \hline
  \multirow{2}{*}{Inception} & \multirow{2}{*} {5.12} & 100 & 2048 & 50K & 200 & 2.21 \\ \cline{3-7}
                            & & 10 & 2048 & 50K & 200 & 0.41 \\ \hline
                            
 \multirow{2}{*}{ResNet-50 \& ResNet-152} & \multirow{2}{*} {5.12} & 100 & 2048 & 50K & 200 & 2.21 \\ \cline{3-7}
                            & & 10 & 2048 & 50K & 200 & 0.41 \\ \hline

 \multirow{2}{*}{EfficientNet} & \multirow{2}{*} {3.2} & 100 & 1280 & 50K & 200 & 2.13 \\ \cline{3-7}
                            & & 10 & 1280 & 50K & 200 & 0.33 \\ \hline
 \multirow{2}{*}{DenseNet} & \multirow{2}{*} {2.56} & 100 & 1024 & 50K & 200 & 2.11 \\ \cline{3-7}
                            & & 10 & 1024 & 50K & 200 & 0.3 \\ \hline                            
                            
 \multirow{2}{*}{MobileNet} & \multirow{2}{*} {2.4} & 100 & 960 & 50K & 200 & 2.1 \\ \cline{3-7}
                            & & 10 & 960 & 50K & 200 & 0.29 \\ \hline
 \multirow{2}{*}{VGG-16} & \multirow{2}{*} {\textbf{1.28}} & 100 & 512 & 50K & 200 & 2.06 \\ \cline{3-7}
                            & & 10 & 512 & 50K & 200 & \textbf{0.25} \\ \hline                   
\end{tabular}\vspace{0pt}
\end{table}

\begin{table}[ht]\vspace{0pt}
\caption{D-SELD Execution Time: LCA encoding stands to benefit from deployment on neuromorphic chips.}
\label{inference-time}
\centering
\begin{tabular}{|l|>{\centering\arraybackslash}m{1.25cm}|>{\centering\arraybackslash}m{2.25cm}|>{\centering\arraybackslash}m{2cm}|>{\centering\arraybackslash}m{2cm}|>{\centering\arraybackslash}m{1.75cm}|}
\hline
 \textbf{Processing Device} & \textbf{Batch Size} & \textbf{Feature Extraction Time (ms)} & \textbf{LCA Encoding Time (ms)} & \textbf{Decoding Time (ms)} & \textbf{Total Time (ms)}\\ \hline
NVIDIA A100 & 64 & 0.9 & 2.52 & 0.1 & 3.52\\ \hline
Apple M1 Pro & 1 & 138&  4689.7 &  2 & 4829.7\\ \hline
\end{tabular}\vspace{-4pt}
\end{table}

\subsection{Comparison to state-of-the-art}
\label{SOTA} 
\begin{table}[t]
\caption{Benchmarking results}
\label{benchmark}
\begin{center}\small
\begin{tabular}{|c|c|c|c|}
        \hline
        \textbf {Dataset} & \textbf {Method} & \textbf {Architecture} & \textbf {Accuracy} \\
        \hline
        \multirow{5}{*}{\textbf {CIFAR-10}} & \textbf{IM-Loss}~\cite{guo2022loss} &  ResNet-19 & \textbf{95.49\%} \\
        \cline{2-4}
        & Dspike~\cite{li2021differentiable}& ResNet-18& 94.25\% \\
        \cline{2-4}
        & OTTT~\cite{xiao2022online}& VGG & 93.73\% \\
        \cline{2-4}
        & Inference on TrueNorth~\cite{TrueNorth-CIFAR} & CNN & 89.32\% \\
        \cline{2-4}
        & \multirow{2}{*}{D-SELD}
        & D-SELD (ResNet-50-V2) & 92.00\% \\
        \cline{3-4}
        & & D-SELD (Resnet-152-V2) & 94.98\% \\
        \hline\hline
        \multirow{6}{*}{\textbf {CIFAR-100}} & Dspike~\cite{li2021differentiable}&  ResNet-18 & 74.24\% \\ 
        \cline{2-4}
        & OTTT~\cite{xiao2022online}& VGG & 71.05\%\\
        \cline{2-4}
        & IM-Loss~\cite{guo2022loss}& VGG-16 & 70.18\% \\
        \cline{2-4}
        & Inference on TrueNorth~\cite{TrueNorth-CIFAR} & CNN & 65.48\% \\
        \cline{2-4}
        & \multirow{3}{*}{\textbf{D-SELD}}
        &D-SELD (ResNet-50-V2) & 74.98\% \\
        \cline{3-4}
        &  & D-SELD (EfficientNet-V2-S) & 75.41\% \\
        \cline{3-4}
        & & \textbf{D-SELD (ResNet-152-V2)} & \textbf{79.32\%} \\
        \hline\hline
        \multirow{6}{*}{\textbf{ImageNet}} & RMP-SNN~\cite{han2020rmp} &   VGG-16& 73.09\% \\
        \cline{2-4}
        & S-ResNet~\cite{9597475} &  ResNet-50& 72.75\% \\
        \cline{2-4}
        & Dspike~\cite{li2021differentiable}&  VGG-16& 71.24\% \\   
        \cline{2-4}
        \cline{2-4}
        & \multirow{3}{*}{\textbf{D-SELD}}&
        D-SELD (ResNet-50-V2) & 76.64\% \\
        \cline{3-4}
        & \textbf{} & D-SELD (EfficientNet-V2-S) & 80.68\% \\
        \cline{3-4}
        & & \textbf{D-SELD (ResNet-152-V2)} & \textbf{80.75\%} \\
        \hline
\end{tabular}
\end{center}
\end{table}
In comparing our results to the state-of-the-art, several critical considerations were addressed to ensure a fair assesment. Firstly, we examined SNNs with neurons that exhibit a (leaky) integrate-and-fire mechanism, considering their suitability for deployment on neuromorphic platforms. 

Secondly, we propose a solution that holds a distinct advantage: it requires no explicit training when utilizing ``Maximum Activation Code" and ''Maximum Sum of Activation Codes" decoders. Deploying D-SELD on Neuromorphic chips entails programming synaptic weights on neural cores using values derived from the dictionary. The only requirement is the construction of the dictionary using extracted feature maps, a process that can be executed off-chip. 

Thirdly, it is important to note that the models we utilized for feature extraction were the original models trained on ImageNet, and we have specified the exact versions of these models used. To maintain the simplicity of D-SELD, we opted not to fine-tune these architectures for specific targeted datasets. The only pre-processing steps applied to the data included resizing, cropping, rescaling, and normalizing. However, should new or improved models become available in the future, their extracted features can still be integrated into D-SELD to potentially enhance its performance further. 

Lastly, in our benchmarking (see Table~\ref{benchmark}), we specifically selected studies  that used SNNs and achieved the highest reported accuracy on all the datasets we tested. Our results indicate that we achieved the highest reported top-1 accuracy of 80.75\% using SNNs on ImageNet and 79.32\% on CIFAR100. For CIFAR10, we attained the second-highest reported accuracy of 94.98\%, with the added benefit of not employing backpropagation and enabling direct deployment on neuromorphic platforms. 

Our benchmarking analysis included results from deploying a CNN model on IBM TrueNorth for CIFAR-10 and CIFAR-100 datasets~\cite{TrueNorth-CIFAR}. However, that study did not include any test results on ImageNet, so direct comparison on that dataset is not performed.

\section{Conclusion}
\label{sec:conclusion}
In this study, we introduce the Dataset-Scalable Exemplar LCA-Decoder approach designed for effective deployment of SNNs on neuromorphic platforms. Our method achieves the highest test accuracy on ImageNet and CIFAR100 datasets, surpassing previous benchmarks for SNN performance on these datasets,
while reducing computational demands associated with conventional error backpropagation methods for SNN training. Moreover, the D-SELD approach offers model scalability, enabling network adjustment to various dataset sizes and incorporation of new data points. This scalability is particularly advantageous in dynamic environments where continuous updates to the training set are necessary without incurring prohibitive training costs. Finally, computational analysis using FLOPs characterizes the training/inference costs associated with our method, demonstrating compelling trade-offs between workload, latency, and performance compared to alternative SNN approaches.


\section*{References}

{
    \small
    \bibliographystyle{unsrt.bst}
    \bibliography{SELD}
}

\end{document}